\documentclass{sigchi-ext}

\usepackage[T1]{fontenc}
\usepackage{textcomp}
\usepackage[scaled=.92]{helvet}
\usepackage{graphicx} 
\usepackage{balance}  
\usepackage{booktabs} 
\usepackage{ccicons}  
\usepackage{ragged2e} 
\usepackage{booktabs}
\usepackage{enumitem}
\usepackage{hyperref}

\def\plaintitle{Towards Using Voice for Hedonic Shopping Motivations} 
\def\emptyauthor{}
\def\plainkeywords{Voice interaction; smart speakers; online shopping; e-commerce; hedonic shopping.}

\title{Towards Using Voice for Hedonic Shopping Motivations}

\numberofauthors{1}
\author{%
  \alignauthor{%
    \textbf{Morteza Behrooz}\\
    \affaddr{University of California, Santa Cruz} \\
    \affaddr{Santa Cruz, CA 95064, USA} \\
    \email{morteza@ucsc.edu} }\alignauthor{%
    \textbf{Preetham Kolari}\\
    \affaddr{eBay}\\
    \affaddr{2025 Hamilton Ave}\\
    \affaddr{San Jose, CA 95125}\\
    \email{pkolari@ebay.com} } \vfil \alignauthor{%
    \textbf{Fred Zaw}\\
    \email{fzaw@gmail.com} }\alignauthor{%
    \textbf{Lindsay Kenzig}\\
    \email{lindsay.kenzig@gmail.com} } \vfil \alignauthor{%
    \textbf{Arnav Jhala}\\
    \affaddr{North Carolina State University}\\
    \affaddr{Raleigh, North Carolina, 27695}\\
    \email{ahjhala@ncsu.com} } 
    }
    
\definecolor{linkColor}{RGB}{6,125,233}
\hypersetup{%
  pdftitle={\plaintitle},
  pdfauthor={\emptyauthor},
  pdfkeywords={\plainkeywords},
  bookmarksnumbered,
  pdfstartview={FitH},
  colorlinks,
  citecolor=black,
  filecolor=black,
  linkcolor=black,
  urlcolor=linkColor,
  breaklinks=true,
}

\begin{document}

\CopyrightYear{2020}
\setcopyright{rightsretained}
\conferenceinfo{IUI'20,}{March  17--20, 2020, Cagliary, Italy}
\isbn{XXX}
\doi{https://doi.org/10.1145/3334480.XXXXXXX}

\copyrightinfo{\acmcopyright}

\maketitle

\RaggedRight{} 

\begin{abstract}
Besides the utilitarian aspects of online shopping, hedonic motivations play a significant role in shaping the shopping behavior of online users \cite{hedonicandutil1}. With the increased popularity of voice-enabled devices \cite{emarketer}, online shopping platforms have attempted to drive online shopping on voice. However, we explain why voice might be more suitable for the hedonic aspects of shopping. We introduce a prototype that enables such focus in a voice experience and share our findings from a qualitative study. 
\end{abstract}

\keywords{\plainkeywords}

\ccsdesc[500]{Human-centered computing~Interaction paradigms}
\ccsdesc[500]{Human-centered computing~Empirical studies in HCI}

\printccsdesc

\section{Motivation and Related Work}
\label{intro}

Today, online retail is a thriving market. In 2017, e-commerce sales amounted to 2.3 trillion USD worldwide, and they are projected to be over twice that amount in 2021 \cite{big-dollars}. Online shopping is also a commonplace user behavior. An estimate of 1.6 billion users around the world purchased goods online in 2017 \cite{many-people}, and 77\% of Internet users in the U.S. (representing 67\% of the population), purchased products online in 2016 \cite{online-shopping-US}. 

In such a large and growing market, recognizing various shopping motivations is crucial to both sales growth and consumer satisfaction. These motivations are usually recognized in terms of being either ``goal-oriented'' and ``utilitarian'', or ``recreational'' and ``hedonic'' \cite{motivations1,motivations2}. Utilitarian shopping is task-oriented and is based on efficiency and rationality, while hedonic shopping is more experiential and is based on curiosity and pleasure \cite{scarpi2012work}. 

\marginpar{%
\vspace{-95pt} \fbox{%
\begin{minipage}{0.925\marginparwidth}
  \textbf{Why is this paper a fit for the IUI 2020 Conversational User Interfaces workshop?} \\ \vspace{6mm}
  
  Dear program committee and reviewers of the workshop, \\ \vspace{6mm}
  
  Voice interaction is widespread, but its user experience paradigms are still often generic and not contextual. In our work, we focus on online shopping, a use case where voice interaction has not been particularly popular, and investigate if a more contextual paradigm of experience is a remedy.  \\ \vspace{3mm}
  
  We, the authors, believe this work and paper to be related to the workshop due to its attempt to expand the use cases of voice interaction, and because of providing an UX paradigm, methodology, and design recommendations for a more contextual experience. 
  \\ \vspace{3mm}
  
  Thank you for your time and consideration.
\end{minipage}} }

Multiple factors such as convenience, time efficiency, availability of aggregate information (e.g. product reviews), and the interactive nature of the involved interfaces have highlighted the utilitarianism of the online shopping experience \cite{hedonicandutil1}; however, hedonism is not exclusive to the brick and mortar retail and is reported to be present in online shopping as well \cite{parsons,funorshopping,hedonicandutil1,hedonicandutil2}. 

\subsection{Hedonic Shopping Motivations}

Hedonic shopping motivations were first introduced as ``non-functional'' motivations \cite{tauber1972people}, and have been regarded as experiential and emotional \cite{hedonicandutil1,funorshopping}. The existence of distinct hedonic motivations of shopping has been confirmed in prior research \cite{hedonictabagh1,hedonicandutil2,flowguy}. In \cite{hedonictabagh2}, for instance, six such motivations are introduced as follows: \textit{Adventure shopping} refers to shopping for stimulation and ``being in another world''. \textit{Social shopping} is the enjoyment of shopping when done with family and friends. \textit{Gratification shopping} is about the stress-relief resulted from shopping. \textit{Role shopping} is buying items for others; e.g. as gifts. \textit{Idea shopping} refers to keeping up with trends, new fashions, and innovations; and finally, \textit{Value shopping} is about shopping for sales and finding bargains.

Even more hedonic shopping motivations have been introduced in prior research. Some of them such as \textit{status} \cite{hedonictabagh1}, which refers to the sense of superiority in receiving service from others, are less applicable to the current forms of online shopping, while other motivations such as online \textit{privacy} are exclusive to online shopping \cite{flowguy}. As far as thirteen distinct hedonic motivations of online shopping are found in a focus group study in \cite{flowguy}, and the ones that are most frequently discussed by the participants in that study are \textit{bargaining} (value shopping), \textit{privacy}, \textit{social}, and \textit{learning trends} (idea shopping).

While some hedonic motivations, such as \textit{value} or \textit{gratification} shopping, are automatically achieved as an inherent part of the act of shopping, others might need or benefit from more explicit facilitation to meet user's needs. For instance, \textit{role} shopping can happen without any explicit support from an online shopping platform, but it can also benefit from a ``gift finding guide''. Similarly, a need for \textit{adventure} in shopping can be met with extensive web browsing sessions, or can be supported explicitly by novel applications such as a Virtual Reality Mall \cite{vrshopp}. As yet another example, online services can facilitate the \textit{social} aspects of shopping by providing information about what one's friends like and buy. The popular social network, Instagram, has recently introduced features that move in this direction \cite{insta}.

In this paper, we focus on \textit{idea shopping} (or \textit{learning trends}). This hedonic motivation is highlighted by many researchers \cite{tauber1972people,hedonictabagh2,flowguy} and is grounded in McGuire's categorization theories \cite{ideaground1}, which explain the human need for structure, order, and knowledge, as well as Festinger's objectification theories \cite{ideaground2}, which explain one's need for external information to make sense of herself. In the absence of explicit facilitation for this hedonic motivation, some users browse online shopping websites for long periods to learn about current trends while there is not necessarily a goal of making a purchase \cite{broooowse}. This behavior is also linked to experiencing positive affect \cite{funsearch}. Our work focuses on more explicit facilitation of this hedonic motivation through voice interfaces, and it would be interesting to see if similar effects are present in the absence of visual stimuli. 

\subsection{Voice Interfaces and Shopping}

The popularity of voice-enabled devices has recently seen a significant surge. According to research conducted by eMarketer \cite{emarketer}, the adoption rate of ``smart speakers'' is only second to the emergence of smartphones in recent years of technology trends and the usage of these devices has surpassed that of wearables \cite{nowear}. It is projected that by 2020, more than 76 million people in the U.S. population (55\% of U.S. households) will use a smart speaker device \cite{emarketer} and more than half of all web searches will be done through voice \cite{search}. 

According to a study in \cite{alexause}, user interactions with smart speakers are most frequently related to music, smart home controls, and general knowledge (e.g. weather). Purchasing, in this study, is reported to only form 0.3\% of user commands. According to another study reported in \cite{sevenlonely}, only 7\% of smart speaker owners have \textit{ever} used these devices to make a purchase. Amazon's Alexa voice assistant, employed in the ``Echo'' smart speakers, is reportedly \cite{alexaftw} the most widely used in the market, but it may not be actively used for shopping \cite{alexahehe}. While there are predictions that suggest more users will make purchases through voice interfaces in the future \cite{clark}, given the current trends discussed above, it is not clear if such forecasts are taking place. 

\begin{marginfigure}[-15pc]
  \begin{minipage}{\marginparwidth}
    \centering
    \includegraphics[width=1\marginparwidth]{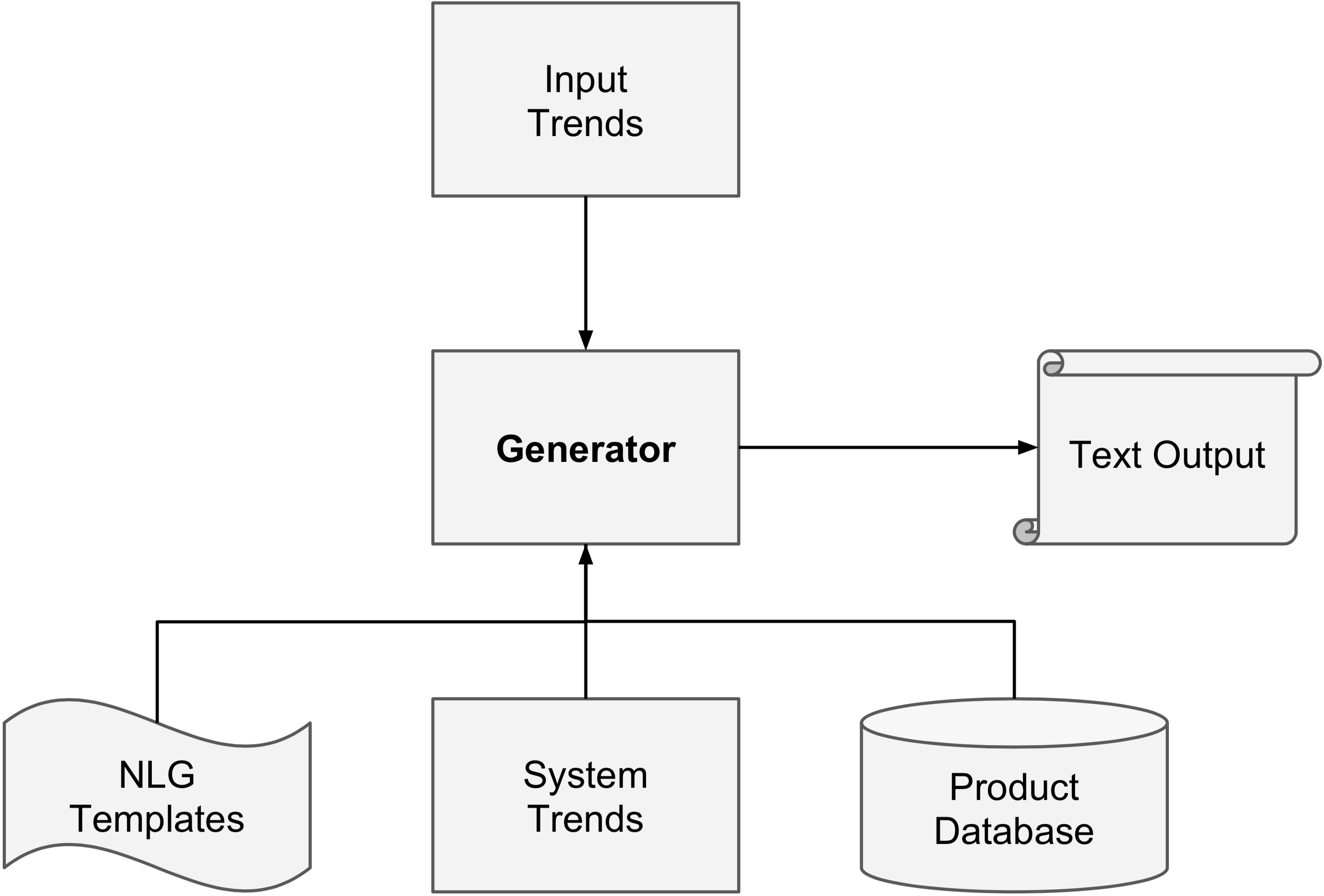}
    \caption{Prototype system's architecture.}~\label{fig:proto}
  \end{minipage}
\end{marginfigure}

We believe that the conversational nature and the human-likeness of the voice interface make it a more appropriate interface for a focus on the hedonic motivations of shopping. In this paper, we describe a simple generator that provides users with sales trends and background product information for a given category of products. To this end, we needed to investigate the benefits and limitations of providing this hedonic value to the users through voice. Our second research question was about how the users might use such experience, both in the context of their daily lives and activities and in terms of its possible effect on their shopping behavior. We hope to inform the content selection for creating such experience and derive design recommendations. Our qualitative user study, aimed in that direction, is described in later sections and followed by a reporting and short embedded discussion of its results. 

\section{Prototype System}

To generate a short voice output about the current sales trends in a given category of product, we developed a prototype that takes as input a set of digested statistics about the sales and search trends, chooses what content to focus on, and outputs readable text describing the chosen trends. The system architecture can be seen in Fig.~\ref{fig:proto}. \vspace{-2mm}


The \textit{input trends} are retrieved from a popular online shopping platform and consist of data about the sales and searches of products. These data are digested and analyzed, previous to being used in our prototype, and can describe comparative statistics across time periods or products. For instance, input trends can point out a surge in the popularity of an item in terms of the number of times that users have searched for it in a given period of time. Other examples include a list of the most popular items, in terms of sales, in a given product category, or a significant change of price for a product (e.g. sales or discounts). \textit{System trends} are a set of common input trends for which we have authored Natural Language Generation templates. 

Some of the corresponding system trends for the example input trends mentioned above are: {\tt ProductPopularitySurge}, {\tt MostPopularProductInCategory}, {\tt ProductDiscountTrend}. Trends that are about a particular product are called \textit{product trends}, and those that are more broadly about a category of product, are called \textit{category trends}. These system trends each have qualifiers that determine the values for their associated variables, such as product IDs, time frames, or discount amounts. \textit{Product database} has a list of products, their categories, brands, current price, and other relevant metadata, including a manually authored ``design story'' which highlights a background fact about a given product.

Given a product category and the input trends for that category, the generator will first look for known system trends among the input trends. Then among the matches, the system looks for category trends and picks one at random. This category trend is used as the first part of the output. To form the second part, the system then focuses on product trends and finds the product that is most frequently the subject of trends (if equal, one at random). This product will become the \textit{focus product} of the generation. Two product trends about the focus product are then chosen to follow the category trend. If no product has two or more trends associated with it, the two trends will be about different products. 

Lastly, these 3 trends are realized into text using NLG templates that are associated with every system trend and the data retrieved from the product database; a design story is also incorporated after the first product trend for the focus product (if there is one). The side note~\ref{tab:output} shows an example output of the system about Sneakers. 


\marginpar{%
  \vspace{-250pt} \fbox{%
    \begin{minipage}{0.925\marginparwidth}
      \textbf{Sample output of the system about Sneakers} \\ \vspace{6mm}
      \small More sneakers dropped recently including Yeezy Boost 700 and Adidas Desert Rat Black ({\tt NewCategoryProductsTrend}). The Adidas Desert Rat Black is the most trending Sneaker ({\tt MostPopularProduct- InCategory}). Not just another basic black sneaker, the latest drop from Yeezy is a tonal mix of black mesh, black suede, and a black retro futuristic 1990s-inspired sole ({\tt Design Story}). The popularity of Adidas Desert Rat Black has increased 30 percent since last month ({\tt ProductPopularitySurge}).
    \end{minipage}}\label{tab:output} }

\section{Evaluation}

To attempt answering our research questions (see Sec.~\ref{intro}) and evaluate our approach in building the prototype, we conducted a semi-structured qualitative user study. We recruited 9 subjects who were non-technical employees of eBay, including 5 females and 4 males; their average age was 29 (min: 21, max: 44).

\subsection{Setup and Procedure}

We used three product categories of sneakers, handbags, and drones to generate three sample output of our generator. To present these outputs to the participants, we used a platform called DialogFlow \footnote{https://dialogflow.com/} which is capable of deploying voice experiences on a Google Home smart speaker. The experience can then be invoked via an activation phrase, e.g. ``demo number 3''. Our DialogFlow experience included a short exchange for the user and started by asking them if they want to learn about what the agent can do. If answered positively by the user, the agent would give a short description of what it is capable of (``I can give you rundowns about what people are searching for, what they are buying, and what is popular in general''). Afterward, the agent offered the user to hear a sample, upon which, the agent read the prototype output from one of the three product categories. One product category was randomly assigned to each of our 9 participants, and each category was used for 3 participants. 

\subsection{Results and Discussion}

In this section, we will lay out the results of our study and discuss them in the context of our research questions.

\subsubsection{\textbf{Participants who already do recreational browsing saw voice as an appropriate medium for receiving trend information}} Out of 9 participants, 4 of them (P1, P7, P8, and P9; all above 30 years of age) declared their shopping behavior to be largely need-based, while the rest of our participants (all below 30 years of age) said they more frequently browse shopping items for fun and look for trend information. The latter group mentioned web search and social media (e.g. Instagram) as their main source of receiving trend information. P3 said: ``\textit{online, I go on Instagram a lot, I follow some influencers and look at where they buy their stuff;}'' and P6, who browses items on a daily basis, said: ``\textit{I go on Google and look at the trends and price information [...] I'm also part of a Facebook group}.'' The same participant later said about the voice demo: ``\textit{this can replace some of the browsing I do every day,}'' and P4 said: ``\textit{[...] what are some fashion trends now? These are questions I like to know the answer to [...] I would be very interested to just use a Google home for this.}''

\subsubsection{\textbf{Need-based shoppers see the value of the voice experience in doing research before buying}} Previously mentioned participants who are need-based shoppers saw the voice experience as a way to research a specific category of product that they have already decided to buy, by comparing prices and seeing what is popular. P7 said: ``\textit{I would use this for bigger things where you spend more money because I would research them more [...] I can get the information and just make the decision later whether I want to buy this or that}.'' P8 and P9 also pointed to such comparative analyses before making a purchase.

\subsubsection{\textbf{This experience can support future purchases and potentially change user's shopping behavior}} The behavior of need-based shoppers does not seem to be affected by the voice experience beyond helping with the research phase of buying, as mentioned in the previous finding. P1 said: ``\textit{it does not change the way I shop [...] if I need to buy something I buy it [...] I'm not a window shopper.}'' Other participants, however, who were not purely need-based shoppers, believed having access to this experience to have the potential to change their shopping behavior. P2, for instance, said: ``\textit{it'd definitely get me on my phone to start searching for stuff.}'' Moreover, multiple participants pointed out a need for finding the items that they would hear about in the voice experience, describing such functionality as ``bookmarking'' (P6) or adding items to a ``wish list'' (P5) as a part of the voice interaction. Giving the users this ability also enables a potential visual experience to follow the main voice experience at a later time, which was noted as something desirable by 2 of our participants (P2 and P4).

\subsubsection{\textbf{Background information about products is desired by trend shoppers}} Participants who were not need-based shoppers appreciated the background information about products, describing it as interesting, while participants who were need-based shoppers did not care about such information. P3, talking about the design story of a handbag product which included information about a celebrity, said: ``\textit{I think the celebrity thing was cool when [the voice] was mentioning celebrities who like [the item].}'' Meanwhile, P8 said of the same information: ``I mean I don't really care personally [...] I'm not super brand-driven.'' 

\subsubsection{\textbf{Voice experience can benefit from including other hedonic motivations}} Even though we did not explicitly ask our participants about other hedonic motivations of shopping (see Sec.~\ref{intro}), they expressed their interest in receiving information about deals and discounts (value shopping), gift guides (role shopping), and shopping as a shared activity (social shopping). P6 said: ``\textit{I'd want more pricing [information], like trending low and trending high, or average maybe.}'' P5 mentioned: ``\textit{when it transitioned to price drop it kind of got my attention.}'' P8, talking about how they would use such voice experience, said: ``\textit{if I was trying to buy someone a gift [...] and it was an area where I wasn't comfortable with, I could see myself [using] that information.}'' And lastly, P9, whom we identified as a need-based shopper, saw this experience as an opportunity for having a shared activity at home: ``\textit{let's say as a family we're just sitting around, we don't have to look at something all of us at the same screen [...], we can just keep asking [via voice] and each one of us can take turns.}''

\subsubsection{\textbf{Voice can introduce shopping-related experiences to new contexts of usage}} Our participants pictured themselves using this voice experience while doing activities that are not usually associated with shopping, such as ``in the car'' (P2), ``in my morning news listening routine'' (P5), ``cooking'' (P7) or ``walking in the backyard'' (P7). 

It would be interesting to study how the voice experience can change and improve based on known user activity (e.g., based on speaker's location), especially to avoid cognitive overload in the user as a result of multitasking.

\subsubsection{\textbf{Users strongly prefer to have an interactive experience}} Participants wanted the experience to be more interactive to control the flow of information and to guide and specify it easier. P4, for example, said: ``\textit{a more structured rundown [would be better:] `hey these are the most popular things, let me know if you want to hear more about a specific shoe' [...] then I can say `tell me more about the Yeezys'.}'' P8, who is a more need-based shopper, wanted even more control and said: ``I would want to be able to ask the specific question that I want the answer to rather than [...] getting a list of information, a more targeted question''. 

As a need-based shopper, P8's desire for detailed question answering is compatible with a tendency to use the experience for researching before buying. Hence, it would be interesting to predict what styles of interactivity might improve a given user's experience on that basis.

\subsubsection{\textbf{The topics covered in the experience can be expanded}} Participants mentioned other topics than trends that they would want to hear about in such experience, including ``news about shops and brands'' (P4) and `'keywords of product reviews'' (P6). Interestingly, P2 mentioned that they would use this experience for content such as ``documentaries'' too.

\section{Conclusion}
Online shopping and usage of voice-enabled devices are both widespread, but it is not clear if buying through voice is becoming popular. In this paper, we introduce a voice prototype that focuses on the hedonic aspects of shopping by providing trend information to the users. We share the results of our qualitative evaluation of this approach and prototype, in which for different types of shoppers, the benefit of providing hedonic values through voice is underlined, its effects on shopping behavior are discussed, and recommendations are made about how such an experience should be designed. 

\balance{} 

\bibliographystyle{SIGCHI-Reference-Format}
\bibliography{bibliography}

\end{document}